\documentclass[preprint,12pt,nofootinbib,groupedaddress,superscriptaddress]{revtex4-2}

\usepackage{graphicx}
\usepackage{dcolumn}
\usepackage{bm}
\usepackage{amssymb}
\usepackage{amsmath}
\usepackage{epsfig}    
\usepackage{color}
\usepackage{slashed}
\usepackage{hyperref}
\usepackage{float}

\newcommand{\green}[1]{{\color[cmyk]{0.97,0,0.75,0}#1}}

\newcommand{\nn}{\nonumber\\}

\newcommand{\be}{\begin{equation}}
\newcommand{\e}{\end{equation}}
\newcommand{\aln}[1]{\begin{align}#1\end{align}}

\newcommand{\R}{{\text{R}}}
\renewcommand{\L}{{\text{L}}}
\newcommand{\I}{{\text{I}}}

\begin{document} 

 
\title{Higgs Alignment from Multicritical-Point Principle \\ in Two Higgs Doublet Models}

\preprint{OU-HET-1229}
\author{Hikaru Kawai}
\email{hikarukawai@phys.ntu.edu.tw}
\affiliation{Department of Physics and Center for Theoretical Physics,  National Taiwan University, Taipei 106, Taiwan}
\affiliation{Physics Division, National Center for Theoretical Sciences, Taipei 10617, Taiwan }
\author{Kiyoharu Kawana}
\email{kkiyoharu@kias.re.kr}
\affiliation{School of Physics, KIAS, Seoul 02455, Korea}
\author{Kin-ya Oda}
\email{odakin@lab.twcu.ac.jp}
\affiliation{Department of Information and Mathematical Sciences, Tokyo Woman’s Christian University, Tokyo 167-8585, Japan}
\author{Kei Yagyu}
\email{yagyu@het.phys.sci.osaka-u.ac.jp}
\affiliation{Department of Physics, Osaka University, Toyonaka, Osaka 560-0043, Japan}



\begin{abstract}\noindent
In models with non-minimal Higgs sectors, enforcing (near) Higgs alignment, necessary to prevent significant deviations in the Higgs boson coupling from the standard model prediction, causes a serious fine-tuning problem. We demonstrate that the Higgs alignment is naturally deduced from the multicritical point principle (MPP) in the general two Higgs doublet model. Furthermore, we discuss the possibility of 
realizing the Yukawa alignment from the MPP, which is necessary to prevent flavor-changing neutral currents mediated by Higgs bosons at tree level.
\end{abstract}
	\maketitle

\newpage
        \tableofcontents

\newpage
\section{Introduction}
Current LHC data reveal two crucial insights: the absence of clear evidence for particles beyond the Standard Model (SM) and the consistent nature of the observed Higgs boson's properties within experimental uncertainties~\cite{ATLAS:2022vkf,CMS:2022dwd}.
This situation is economically explained by the SM whose Higgs sector is a minimal form composed of a single isospin doublet field. 
However, new physics beyond the SM must exist because the SM cannot explain the established phenomena such as neutrino oscillations, existence of dark matter and baryon asymmetry of the Universe.  
This leads to the perplexing question of why signs of new physics have not been observed despite their necessity.

One simple explanation is that the scale of new physics is much greater than the electroweak scale, leading to a decoupling of its effects from the SM. 
The canonical seesaw scenario~\cite{Minkowski:1977sc} is one of the well known examples of such new physics, in which super heavy right-handed neutrinos are introduced to explain neutrino masses and oscillations~\cite{Yanagida:1979as,Gell-Mann:1979vob,Mohapatra:1979ia} as well as baryon asymmetry via the leptogenesis~\cite{Fukugita:1986hr}.
Such scenarios, if realized, are generally challenging to test experimentally.

On the other hand, there are possibilities of realizing new physics at the electroweak and/or TeV scales. In these scenarios, the Higgs sector is often extended from the minimal form. In such models with non-minimal Higgs sectors, the couplings of the SM-like Higgs boson, which can be identified with one of the neutral Higgs bosons, can deviate from the SM predictions due to scalar mixings. 
%
Therefore, the so-called Higgs alignment~\cite{Gunion:2002zf} is required, that is the assumption that the mass eigenstate of the Higgs boson identified with the observed one does not mix with the other neutral states. 
%

The Higgs alignment can naturally be realized by imposing an unbroken $\mathbb{Z}_2^{}$ symmetry such as the inert doublet model~\cite{Barbieri:2006dq}, but it spoils the CP-violating phases which are required for the successful electroweak baryogenesis, see e.g., Refs.~\cite{Turok:1990in,Turok:1990zg,Cline:1995dg} as earlier works in two Higgs doublet models (2HDMs). 
 We thus encounter a fine-tuning problem when we need to impose the Higgs alinment in the non-minimal Higgs sectors.
 
The Multicritical-Point Principle (MPP)~\cite{Bennett:1993pj,Froggatt:1995rt,Nielsen:2012pu} offers a natural solution to such fine-tuning issues. The MPP can be described in such a way that coupling constants in a theory are naturally tuned to one of the critical points at which quantum phase transitions occur. We briefly outline the essence of this mechanism; see Ref.~\cite{Kawai:2023viy} for a more detailed account. 

Consider a parameter $\lambda$, such as a mass or a coupling constant, in the action $S_\lambda^{}[\phi]$. In conventional quantum field theory (QFT), a canonical partition function is used:
\aln{Z_{\rm can}^{}(\lambda)=\int{\cal D}\phi \exp\left(i\left\{S_0^{}[\phi]+\lambda S_\lambda^{}[\phi]\right\}\right),}
where $S_0^{}[\phi]$ is the other parts irrelevant to $\lambda$. Here, $\lambda$ is a free parameter with no inherent principle for specific values, leading to the fine-tuning problem. Alternatively, a micro-canonical formulation fixes $S_\lambda^{}[\phi]$ to a certain value $A$:
\aln{Z_{\rm mic}^{}(A)=\int {\cal D}\phi e^{iS_0^{}}\delta\left(S_\lambda^{}[\phi]-A\right)=\int_{-\infty}^{\infty}\frac{d\lambda}{2\pi}e^{i\lambda A}Z_{\rm can}^{}(\lambda).}
If the integration over $\lambda$ is dominated by a point $\lambda=\lambda_*^{}(A)$ as $V\rightarrow \infty$, the micro-canonical formulation effectively becomes a canonical one with $\lambda$ fixed at $\lambda_*^{}(A)$. A simple example is a saddle point where
\aln{e^{i\lambda A}Z_{\rm can}^{}(\lambda)\propto e^{iV\left[(\lambda-\lambda_*^{}(A))^2+\cdots \right]},}
leading to
\aln{\log Z_{\rm mic}^{}(A)=\log Z_{\rm can}^{}(\lambda )\bigg|_{\lambda=\lambda_*^{}(A)}\left(1+{\cal O}(V^{-1})\right)}
for $V\rightarrow \infty$. Refer to Refs.~\cite{Kawai:2021lam,Hamada:2022dan,Kawai:2023viy} for recent developments in this area.

In this paper, we show how the MPP naturally leads to the Higgs alignment in the most general 2HDM without arbitrary $\mathbb{Z}_2$ symmetries. This approach can facilitate successful scenarios of electroweak baryogenesis~\cite{Liu:2011jh,Ahmadvand:2013sna,Chiang:2016vgf,Guo:2016ixx,Fuyuto:2017ewj,Modak:2018csw,Enomoto:2021dkl,Enomoto:2022rrl} due to sufficient CP-violating phases in Yukawa interactions under the constraints by electric dipole moments~\cite{Kanemura:2020ibp,Fuyuto:2019svr}. We also discuss the possibility of realizing the Yukawa alignment~\cite{Pich:2009sp} which is required to avoid flavor-changing neutral currents (FCNCs) at tree level, through the MPP.

The paper is organized as follows. Section~\ref{sec:fine-tuning} reviews the general 2HDM, highlighting fine-tuning issues associated with the Higgs and Yukawa alignments. Section~\ref{sec:mpp} demonstrates the natural derivation of the Higgs alignment via the MPP. 
In Section~\ref{sec:Yukawa-alignment}, we discuss the possibility to realize the Yukawa alignment from the MPP. 
Conclusions are given in Section~\ref{sec:conclusion}. 
In Appendix.~\ref{sec:canonical}, we present how we can obtain the canonical kinetic terms for the physical Higgs boson in the unitary gauge employed in this study.

%

\section{Fine-tuning problem caused by the Alignments \label{sec:fine-tuning}}

We address the fine-tuning problem arising from the Higgs alignment and the Yukawa alignment in the 2HDM. 
We consider the general model without imposing any additional symmetries such as a $\mathbb{Z}_2$ symmetry. 
%

The Higgs sector of the 2HDM is composed of two isospin doublet fields $\tilde{\Phi}_1$ and $\tilde{\Phi}_2$. 
Since these two fields have the same charge and representation, we have degrees of freedom to rotate them as
\begin{equation}
\begin{pmatrix}
\tilde{\Phi}_1 \\
\tilde{\Phi}_2 
\end{pmatrix}
= U 
\begin{pmatrix}
\Phi_1 \\
\Phi_2
\end{pmatrix},\quad U\in U(2). \label{eq:higgs-basis}
\end{equation}
Among the various choices for $(\Phi_1,\Phi_2)$, the Higgs basis~\cite{Davidson:2005cw} is frequently utilized due to its clarity in conveying the physical properties of the scalar fields, as will be demonstrated. 
In order to distinguish the Higgs basis and other bases, we particularly represent the former as $H_1 \equiv \Phi_1$ and $H_2 \equiv \Phi_2$, and they can be parameterized as 
\begin{align}
H_1 = \begin{pmatrix}
G^+ \\
\frac{1}{\sqrt{2}}(h_1' + v + iG^0)
\end{pmatrix},\quad 
H_2 = \begin{pmatrix}
H^+ \\
\frac{1}{\sqrt{2}}(h_2'  + ih_3') 
\end{pmatrix}, \label{eq:higgs}
\end{align}
where $G^\pm$ and $G^0$ are the Nambu-Goldstone (NG) bosons which are absorbed into the longitudinal components of $W^\pm$ and $Z$ bosons, respectively, while 
$H^\pm$ $(h_{1,2,3}')$ are the physical singly-charged (neutral) Higgs bosons. 
In this basis, only $H_1$ obtains the non-zero vacuume expectaion value (VEV) $v$ defined as $v \equiv (\sqrt{2}G_F)^{-1}$ with $G_F$ being the Fermi constant, 
so that $H_1$ behaves as the Higgs field. 
This means that the neutral component $h_1'$ plays a role of the Higgs boson in the SM, because 
the gauge interaction and the Yukawa interaction for $h_1'$ are the same as those of the SM Higgs boson at tree level. 
We here have implicitly assumed that the configuration of the VEVs in Eq.~(\ref{eq:higgs}), i.e., $\langle H_1 \rangle= (0,v/\sqrt{2})^T$ and $\langle H_2 \rangle =0$, corresponds to the global minimum of the potential. 

We note that the \( h_1' \) state generally does not correspond to a mass eigenstate\green{: I}t may mix with the other neutral states, \( h_2' \) and \( h_3' \), as will be discussed in the following subsection. Such mixing can lead to deviations in the couplings of \( h_1' \) from those predicted for the SM Higgs boson. The {\it Higgs alignment limit} is thus the one in which the \( h_1' \) state does not mix with the \( h_2' \) and \( h_3' \) states. In this limit, the couplings of \( h_1' \) align with the SM predictions at the tree level. 

\subsection{Higgs alignment}
The most general form of the Higgs potential is written in the general basis ($\Phi_1$,$\Phi_2$) as  
\aln{V({\Phi}_1^{},{\Phi}_2^{})= & {\mu}_1({\Phi}_1^\dagger {\Phi}_1^{})  +{\mu}_2({\Phi}_2^\dagger {\Phi}_2^{}) + ({\mu}_3 {\Phi}^\dagger_1 {\Phi}_2^{}+ \text{h.c.})
\nn
& + \frac{{\lambda}_1}{2} ({\Phi}_1^{\dagger}{\Phi}_1^{})^2 + \frac{{\lambda}_2}{2} ({\Phi}_2^{\dagger}{\Phi}_2^{})^2 + {\lambda}_3  ({\Phi}_1^{\dagger} {\Phi}_1^{})  ( {\Phi}_2^{\dagger} {\Phi}_2^{}) +  {\lambda}_4  ( {\Phi}_1^{\dagger} {\Phi}_2^{})( {\Phi}_2^{\dagger} {\Phi}_1^{}) 
\notag \\
&+ \left[\frac{ {\lambda}_5}{2}  ( {\Phi}_1^{\dagger} {\Phi}_2^{})^2 +  {\lambda}_6 ( {\Phi}_1^{\dagger} {\Phi}_1^{})( {\Phi}_1^{\dagger} {\Phi}_2^{}) + {\lambda}_7 ( {\Phi}_2^{\dagger} {\Phi}_2^{})( {\Phi}_1^{\dagger}{\Phi}_2^{}) + \text{h.c.} \right], \label{eq:pot}
}
where ${\mu}_3$ and ${\lambda}_{5,6,7}$ are complex parameters, and the others are real. 
%
Here we consider the Higgs basis as introduced in Eq.~(\ref{eq:higgs})  
i.e., $\Phi_{1,2} = H_{1,2}$. 
To clearly indicate the use of the Higgs basis, we re-label the parameters in the Higgs potential from $\mu_i$ to $M_i$ and from $\lambda_j$ to $\Lambda_j$.

The stationary conditions, 
\begin{align}
\frac{\partial V}{\partial h_i^\prime}\Bigg|_{0} = 0~~(i=1,\dots,3),
\end{align}
yield the following relations:  
\begin{align}
M_1 &= -\frac{v^2}{2}\Lambda_1,&
M_3 &= -\frac{v^2}{2}\Lambda_6, \label{eq:tadpole}
\end{align}
where $|_0$ denotes that all fields are set to zero.

By applying Eq.~(\ref{eq:tadpole}), we eliminate the parameters $M_1$ and $M_3$. The mass of the charged Higgs boson is then given by
\begin{align}
  m_{H^\pm}^2 = M_2 + \frac{\Lambda_3}{2}v^2, 
\end{align}
and the mass matrix for the neutral Higgs bosons in the $(h_1',h_2',h_3')$ basis is 
\begin{align}
  \mathcal{M}_{\rm neut}
  =	v^2
    \begin{pmatrix}
      \Lambda_1	& \text{Re}\Lambda_6 & -\text{Im}\Lambda_6 \\
      \text{Re}\Lambda_6 & \frac{M_2}{v^2} + \frac{1}{2}(\Lambda_{3} + \Lambda_{4} + \text{Re}\Lambda_{5})&-\frac{1}{2}\text{Im}\Lambda_5 \\
      -\text{Im}\Lambda_6 &-\frac{1}{2}\text{Im}\Lambda_5 & \frac{M_2}{v^2}+\frac{1}{2}(\Lambda_3+\Lambda_4-\text{Re}\Lambda_5)
    \end{pmatrix}. 
    \label{eq:massmatrix}
    \end{align}
It is evident that the Higgs alignment limit can be expressed by the conditions:
\begin{align}
({\cal M}_{\rm neut})_{12} &\to 0,&
({\cal M}_{\rm neut})_{13} &\to 0. 
\end{align}
This can be realized by $\Lambda_6^{}=0$, which is equivalent to the following condition according to Eq.~(\ref{eq:tadpole}):
\begin{align}
M_3 = 0. \label{eq:higgs-alignment}
\end{align}
%
It is noteworthy that the Higgs alignment naturally occurs in the decoupling limit as $M_2 \to \infty$ because the mixings are suppressed by $v^2/M_2 \to 0$ in this limit.  However, this scenario is less interesting from a phenomenological standpoint since all predictions converge to those of the SM, diminishing the various motivations for considering the 2HDM.

One of the simplest ways to realize the Higgs alignment is by imposing an unbroken \(\mathbb{Z}_2\) symmetry in Eq.~(\ref{eq:pot}), such that \(\Phi_1 \to + \Phi_1\) and \(\Phi_2 \to -\Phi_2\). This symmetry explicitly forbids the terms \(\mu_3^{}\) and \(\lambda_{6,7}\). Consequently, the Higgs alignment is achieved through this symmetry. This approach leads to the so-called inert doublet model~\cite{Barbieri:2006dq}, which is a compelling model for explaining the existence of dark matter. However, it does not support electroweak baryogenesis due to the absence of new CP-violating phases. 
%


As we have seen above, symmetries tend to impose too strict constraints on the parameter space. 
On the other hand, the MPP naturally realizes fine-tuning of parameters without such an over-constraint.    
In the next section, we will show that the MPP can indeed realize the 
conditions for the Higgs alignment without the extra constraints, contrary to the case with additional symmetries. 
%
%
In order to do that, it would be useful to restart from the general basis~(\ref{eq:pot}), instead of the Higgs basis. 
In particular, we can choose the basis with $\lambda_6 = 0$ (see the discussion in Sec.~\ref{sec:mpp}), in which 
the necessary and sufficient condition of the Higgs alignment turn out to be
\begin{align}
\mu_3&= 0,&
\langle \Phi_2 \rangle &= 0. 
\label{eq:higgs-alignment2}
\end{align}


\subsection{Yukawa alignment}

The most general form of the Yukawa interactions are described as follows:
\begin{align}
{\cal L}_Y & = 
-\sum_{a=1}^2\sum_{i,j=1}^3\left[\overline{{Q}_{\L i}^{}}({Y}_d^{(a)})^{ij} {\Phi}_a {d}_{\R j}^{}
+\overline{{Q}_{\L i}}({Y}_u^{(a)})^{ij} {\Phi}^c_a {u}_{\R j}^{}
+\overline{{L}_{\L i}}({Y}_e^{(a)})^{ij}{\Phi}_a {e}_{\R j}^{}
+\text{h.c.}\right],
\label{Eq:Yukawa0}
\end{align}
where ${Y}_f^{(a)}$ ($f=u,d$ and $e$) are $3 \times 3$ complex matrices, and $\Phi^c_a\equiv i\tau_2 \Phi_a^*$. 
Here, ${Q}_{\L i}$ and ${L}_{\L i}$ represent the left-handed quark and lepton doublets of the $i$th generation, respectively, while ${u}_{\R i}$,  ${d}_{\R i}$, and  ${e}_{\R i}$ denote the right-handed up-type quarks, down-type quarks, and charged leptons.
In the following, we suppress the flavor indices for simplicity. 

In the Higgs basis, we identify $\Phi_1^{}$ to $H_1^{}$ and $\Phi_2^{}$ to $H_2^{}$. Utilizing a bi-unitary transformation, $f_\L ^{} \to V_f f_\L ^{}$ and $f_\R ^{} \to U_f f_\R ^{}$ with $V_f^{}, U_f^{}\in U(3)$ and $f_\L ^{}$ and $f_\R ^{}$ representing left-handed and right-handed fermions, respectively,\footnote{
We use the same symbols $f_{\L ,\R}^{}$ for fermions before and after the transformation for simplicity.
}
one of the Yukawa matrices can be diagonalized as
%
\aln{
&Y_{f}^{(1)}\to V_f^{\dagger}\, Y_{f}^{(1)}\, U_f^{} =: \frac{\sqrt{2}}{v}M_f, \\
&Y_{f}^{(2)}\to V_f^{\dagger}\,Y_{f}^{(2)} \, U_f^{} =: \rho_f,
}
where $M_f$ denotes a diagonal mass matrix with real-positive eigenvalues, and $\rho_f$ is a general complex $3\times 3$ matrix. 
%
Consequently, Eq.~(\ref{Eq:Yukawa0}) becomes
\begin{align}
\mathcal{L}_Y^{} = -\Bigg[&\overline{Q_\L ^{}} \left(\frac{\sqrt{2}}{v}M_d H_1^{} + \rho_d H_2\right)d_\R ^{}
+\overline{Q_\L ^{}}V_{\rm CKM}^\dagger\left(\frac{\sqrt{2}}{v}M_u H_1^c + \rho_u H_2^c\right)u_\R ^{} \notag\\
& +\overline{L_\L ^{}}\left(\frac{\sqrt{2}}{v}M_eH_1 + \rho_e H_2\right) e_\R ^{} +\text{h.c.}\Bigg],  \label{BottomUp:2HDM:Yukawa1}
\end{align}
where $V_{\rm CKM}^{}=V_u^\dagger V_d^{}$ is the Cabibbo-Kobayashi-Maskawa (CKM) matrix, and 
\aln{
{Q}_\L ^{}=\begin{pmatrix}V_{\rm CKM}^{\dagger}{u}_\L ^{}
\\d_\L ^{}\end{pmatrix}.
}
It is apparent that off-diagonal elements of $\rho_f$ can lead to FCNCs mediated by the $h_{2,3}'$ and/or $H^\pm$ fields. These FCNCs occur even in the Higgs alignment limit, where the Yukawa couplings for $h_{2,3}$ provide flavor-violating interactions.
%

The Yukawa alignment, originally proposed in Ref.~\cite{Pich:2009sp}, is defined as
\begin{align}
Y_f^{(2)}  \propto  Y_f^{(1)}. \label{eq:yukawa-alignment}
\end{align}
In the Higgs basis, this alignment can be expressed as 
\begin{align}
\rho_f  = \zeta_f \frac{\sqrt{2}M_f}{v},
\label{eq:yukawa-alignment2}
\end{align}
where $\zeta_f$ represents a complex, non-matrix parameter. 
%
However, from a theoretical view point, there is no inherent reason to assume such a Yukawa alignment as defined in (\ref{eq:yukawa-alignment}). This lack of theoretical background further motivates the exploration of the realization of the Yukawa alignment through the MPP, which will be discussed in Sec.~\ref{sec:Yukawa-alignment}.


%
\section{Higgs alignment via MPP \label{sec:mpp}}
We now demonstrate how the Higgs alignment condition, as shown in Eq.~(\ref{eq:higgs-alignment2}), can be achieved using the MPP.

\subsection{$\lambda_6=0$ basis}
In analyzing the vacuum structure of the 2HDM, it is useful to apply the \(U(2)\) transformation~(\ref{eq:higgs-basis}) to simplify the potential~(\ref{eq:pot}) by eliminating some coupling constants.
%

An \(SU(2)\) component of this transformation,
\aln{
U&=e^{ic_1^{}\tau_1^{}+ic_2^{}\tau_2^{}}=\left(\begin{matrix} \cos|c|& \frac{c}{|c|}\sin|c|
\\
-\frac{c^*}{|c|}\sin|c|&  \cos|c|
\end{matrix}
\right),&
c&=c_2^{}+ic_1^{}\quad( c_{1,2}\in \mathbb{R})
\label{SU(2) part}
}
mixes the two Higgs fields.
%
By using this transformation, one can eliminate one of the complex coupling constants $\mu_3$, $\lambda_5^{},\lambda_6^{}$ and $\lambda_7^{}$ in a finite region of the parameter space.  
In the following, we take\footnote{
In fact, one can see that $(\tilde{\Phi}_1^{\dagger}\tilde{\Phi}_1^{})^2$ term gives $\sim c(\Phi_1^\dagger \Phi_1^{})(\Phi_1^{\dagger}\Phi_2^{}+{\rm h.c})$ in the new basis, which can be used to cancel $\lambda_6^{}$ unless it is too large compared with the other coupling constants. This is sufficient because the philosophy of MPP is to realize automatic fine-tuning within a finite region of parameter space. 
}
\begin{align}
\lambda_6^{}=0.
\end{align}
%
Furthermore, we can take $\mu_3^{}\in \mathbb{R}$ by using the remaining phase transformations, i.e. $U={\rm diag}(e^{i\theta_1^{}},e^{i\theta_2^{}})$. 
Finally, we note that this basis does not necessarily coincide with the Higgs basis.

%

\subsection{Taking unitary gauge}
Next, by employing the usual $SU(2)$ gauge invariance, we can choose the unitary gauge as
\aln{
\Phi_1  &=\left(\begin{matrix} 0 \\ 
\frac{\eta}{\sqrt{2}}
\end{matrix}
\right),&
\Phi_2  &=\left(\begin{matrix} \varphi^+ \\ 
\varphi^0
\end{matrix}
\right),&
\eta    &\in \mathbb{R},\quad
\varphi^\pm,\varphi^0\in \mathbb{C}. \label{eq:kawai-basis}
}  
Note that when \(\langle \varphi^\pm \rangle = \langle \varphi^0 \rangle = 0\), the fields set to zero in Eq.~(\ref{eq:kawai-basis}) correspond precisely to the unphysical NG bosons. 
Conversely, if there are non-zero VEVs for \(\Phi_2\), the \(\varphi^\pm\) and \(\varphi^0\) fields can mix with the longitudinal components of the weak gauge bosons via the kinetic term (see Appendix~\ref{sec:canonical} for details ).

The potential, when rewritten in the new basis and in the unitary gauge, is expressed as
\aln{V(\eta,\varphi^\pm,\varphi^0)&=
\frac{\mu_1}{2}\eta^2 + \mu_2\left(\left|\varphi^0\right|^2 + \left|\varphi^+\right|^2\right) + \frac{\mu_3}{\sqrt{2}}\eta\left(\varphi^0 + \varphi^{0*}\right) \notag\\
& + \frac{\lambda_1}{8}\eta^4  +\frac{\lambda_2}{2}\left(\left|\varphi^0\right|^2+\left|\varphi^+\right|^2\right)^2 + \frac{\lambda_3}{2}\eta^2\left(\left|\varphi^0\right|^2+\left|\varphi^+\right|^2\right) +\frac{\lambda_4}{2}\eta^2\left|\varphi^0\right|^2 \notag\\
& + \frac{\lambda_5}{4}\eta^2\left(\varphi^0\right)^2 + \frac{\lambda_7}{\sqrt{2}}\eta \varphi^0\left(\left|\varphi^0\right|^2 + \left|\varphi^+\right|^2\right) + \text{h.c.} 
\label{effective potential 1}
}

\subsection{Charged Higgs VEV}
Let us first examine the VEV for the $\varphi^\pm$ fields. 
Since $\varphi^\pm$ appear only via the $\left|\varphi^+\right|^2$ form in the potential, we can determine whether $\langle \varphi^\pm \rangle = 0$ or $\langle \varphi^\pm \rangle \neq 0$ by looking at the sign of the mass term given as 
\aln{
m_{\varphi^\pm}^2 \equiv \frac{\partial^2 V}{\partial \varphi^+ \partial \varphi^-}\bigg|_{\varphi^\pm = 0}^{}= \mu_2 + \lambda_2 \left|\varphi^0\right|^2 + \frac{\lambda_3}{2}\eta^2 
+ \sqrt{2}\eta\,\text{Re}\left(\lambda_7\varphi^0\right). 
\label{VEV condition}
}
To prevent a charge-breaking vacuum \(\langle \varphi^\pm \rangle \neq 0\), we ensure \(m_{\varphi^\pm}^2 > 0\), imposing an inequality among the coupling constants. 

\subsection{Neutral Higgs VEV}
Next, we examine the VEV for the neutral component $\varphi^0$. This analysis is conducted by setting $\varphi^\pm = 0$ in Eq.~(\ref{effective potential 1}), leading to the following expression for the potential:
\aln{
 V(\eta,\varphi^0)&=
 \frac{\mu_1}{2}\eta^2  + \frac{\lambda_1}{8}\eta^4 + \mu_2 \left|\varphi^0\right|^2   +\frac{\lambda_2}{2}|\varphi^0|^4  \notag\\
 &\quad+ \sqrt{2}\mu_3\eta\,\text{Re}\left(\varphi^0\right) + \frac{1}{2}\left(\lambda_3+\lambda_4\right)\eta^2 \left|\varphi^0\right|^2   + 
 \left[\frac{\lambda_5}{4}\eta^2\left(\varphi^0\right)^2 + \frac{\lambda_7}{\sqrt{2}}\eta \varphi^{0}\left|\varphi^0\right|^2  + \text{h.c.} \right].  
 \label{effective potential 2}
 }
For a given value of $\langle \eta \rangle$, we can express the potential in terms of the real and imaginary parts, $\varphi^0 = (\varphi_\R  + i \varphi_\I )/\sqrt{2}$:
\begin{align}
V(\varphi_\R ^{},\varphi_\I ^{}) & =  
\mu_3\langle\eta\rangle\varphi_\R ^{}  
+ \frac{1}{2}\begin{pmatrix}\varphi_\R&\varphi_\I\end{pmatrix}{\cal M}
\begin{pmatrix}
\varphi_\R \\
\varphi_\I 
\end{pmatrix}+ (\text{cubic and quartic terms of }\varphi_{\R,\I}^{}), \label{eq:pot3}
\end{align}
where ${\cal M}$ is a $2\times 2$ matrix defined by 
\begin{align}
{\cal M} \equiv  
\begin{bmatrix}
\frac{\partial^2 V}{\partial \varphi_\R ^2} & \frac{\partial^2 V}{\partial \varphi_\R \partial \varphi_\I } \\
\frac{\partial^2 V}{\partial \varphi_\R \partial \varphi_\I } & \frac{\partial^2 V}{\partial \varphi_\I ^2} 
\end{bmatrix}_{\varphi_\R ^{} = \varphi_\I ^{} = 0} = 
 \begin{pmatrix}
\mu_2 + \frac{\langle\eta\rangle^2}{2}\left(\lambda_3 + \lambda_4 + \text{Re}\lambda_5\right)  & -\frac{\langle\eta\rangle^2}{2}\text{Im} \lambda_5 \\
-\frac{\langle\eta\rangle^2}{2}\text{Im} \lambda_5 & \mu_2 + \frac{\langle\eta\rangle^2}{2}\left(\lambda_3 + \lambda_4 - \text{Re}\lambda_5\right)
\end{pmatrix}.
\end{align}

\subsection{Vacuum structure when $\mu_3=0$}
Here and in the next subsection, we show that, when regarding the vacuum energy as a function of $\mu_3$, the vacuum energy becomes extremum at the point $\mu_3=0$.
In this sense, $\mu_3=0$ is a critical point that can be chosen by the MPP.

First, we discuss the conditions under which $\varphi_\R=\varphi_\I=0$ is a true vacuum when $\mu_3=0$.
It is clear that in the limit of $\mu_3 \to 0$, the potential~(\ref{eq:pot3}) has a local minimum at $\varphi_\R ^{} = \varphi_\I ^{} = 0$ if 
the following conditions are satisfied: 
\begin{align}
|{\cal M}| &> 0,&  {\cal M}_{22} &> 0, \label{eq:hesse}
\end{align}

Even for $\mu_3^{} = 0$, there can be the other local minima at $\varphi_\R ^{} \neq 0$ and/or  $\varphi_\I ^{} \neq 0$ due to the existence of the cubic terms of $\varphi_\R ^3$ and  $\varphi_\I ^3$. 
In order to examine such a minimum, 
it would be convenient to move to the new basis of ($\varphi_\R ^{}$, $\varphi_\I ^{}$) by which the matrix ${\cal M}$ is diagonalized. Namely, 
\begin{align}
\begin{pmatrix}
\varphi_\R ^{} \\
\varphi_\I ^{}
\end{pmatrix}
=
\begin{pmatrix}
\cos\theta & - \sin\theta \\
\sin\theta & \cos\theta
\end{pmatrix}
\begin{pmatrix}
\varphi_\R ^\prime \\
\varphi_\I ^\prime
\end{pmatrix}, \qquad\text{with}\quad\tan 2\theta = -\frac{\text{Im} \lambda_5}{\text{Re} \lambda_{5}}. 
\end{align}
In this new basis, the potential (\ref{eq:pot3}) with $\mu_3 = 0$ becomes
\aln{
V\!\left(\varphi_\R ^\prime,\varphi_\I ^\prime\right)&=
\frac{m_\R ^2}{2}\varphi_\R ^{\prime 2}
+\frac{m_\I ^2}{2}\varphi_\I ^{\prime 2}
+\left(\kappa_\R ^{}\varphi_\R ^{\prime}+\kappa_\I ^{} \varphi_\I ^{\prime}\right)\left(\varphi_\R ^{\prime 2} + \varphi_\I ^{\prime 2}\right)
+\frac{\lambda_2}{8}\left(\varphi_\R ^{\prime 2} + \varphi_\I ^{\prime 2}\right)^2, \label{eq:pot4}
}
where 
\begin{align}
m_\R ^2 &= \mu_2 +\frac{\langle \eta\rangle^2}{2} \left(\lambda_3 + \lambda_4     + \frac{\text{Re}\lambda_5}{\cos2\theta}\right), \\
m_\I ^2 &= \mu_2 +\frac{\langle \eta\rangle^2}{2} \left(\lambda_3 + \lambda_4     - \frac{\text{Re}\lambda_5}{\cos2\theta}\right), \\
\kappa_\R ^{} &=  \frac{\langle \eta\rangle}{2} \left(\text{Re} \lambda_7 \cos\theta - \text{Im} \lambda_7 \sin\theta\right), \\
\kappa_\I ^{} &=  -\frac{\langle \eta\rangle}{2} \left(\text{Re} \lambda_7 \sin\theta + \text{Im} \lambda_7 \cos\theta\right). 
\end{align}
The condition (\ref{eq:hesse}) is then rewritten as
\begin{align}
m_\R ^2 &> 0,& 
m_\I ^2 &> 0.  \label{eq:condition1}
\end{align}

Now the absence of other local minima is guaranteed when
\begin{align}
\frac{\partial V(\varphi_\R ^\prime,\varphi_\I ^\prime)}{\partial \varphi_\R ^\prime}
&\neq 0,&
\frac{\partial V(\varphi_\R ^\prime,\varphi_\I ^\prime)}{\partial \varphi_\I ^\prime}
&\neq 0,&
\text{for}\quad \varphi_{\R,\I}^{\prime} &\neq 0. 
\label{eq:cond2}
\end{align}
We find that the condition~\eqref{eq:cond2} is satisfied when 
\begin{align}
\frac{\kappa_\R ^2}{m_\R ^2} + \frac{\kappa_\I ^2}{m_\I ^2} < \frac{2}{9}\lambda_2.
\label{eq:condition2}
\end{align}
One can numerically check that the inequality (\ref{eq:condition2}) is the sufficient condition to realize the condition~(\ref{eq:cond2}).  
Note that even if this condition is not satisfied, the minimum at the origin could still be the global minimum, since other potential minima away from the origin may have higher values.

In any case, over a wide range of the parameter space,  $\varphi^{0} =0$ is the global minimum of the potential when $\mu_3 =0$.

\subsection{$\mu_3=0$ is extremum of vacuum energy}
Next, we discuss how the vacuum energy varies as a function of $\mu_3^{}$.
When a small value of $\mu_3^{}$ is turned on, non-zero VEVs $\langle \varphi_\R ^{}\rangle$ and $\langle \varphi_\I ^{}\rangle$ are induced as~\footnote{
It is equivalent to 
\begin{align*}
\langle \varphi_\R ^{\prime}\rangle &= -\frac{\langle\eta\rangle\cos\theta}{m_\R ^2}\mu_3 + {\cal O}\!\left(\mu_3^2\right),&
\langle \varphi_\I ^{\prime}\rangle &= \frac{\langle\eta\rangle\sin\theta}{m_\I ^2}\mu_3 + {\cal O}\!\left(\mu_3^2\right). 
\end{align*}
} 
 \begin{align}
 \langle \varphi_\R ^{}\rangle = -\frac{{\cal M}_{22}}{|{\cal M}|}\langle\eta\rangle\mu_3 + {\cal O}\!\left(\mu_3^2\right),\quad
 \langle \varphi_\I ^{}\rangle = \frac{{\cal M}_{12}}{|{\cal M}|}\langle\eta\rangle\mu_3 + {\cal O}\!\left(\mu_3^2\right). 
 \end{align}
At the same time, the VEV of $\eta$ is slightly shifted to the $\varphi_\R ^{}$ direction as
\begin{align}
\langle \eta \rangle^2 = v_0^2 + \sqrt{\frac{2}{-\lambda_1\mu_1}}\langle \varphi_\R ^{} \rangle\mu_3 + {\cal O}(\mu_3^3), 
\end{align}
where 
\begin{align}
v_0^2 \equiv \langle \eta \rangle^2\Big|_{\mu_3 = 0} = -\frac{2\mu_1}{\lambda_1}.
\end{align}
The vacuum energy ${\cal E}$ with a small $\mu_3$ is then expressed as 
\begin{align}
{\cal E}  = {\cal E}_0 + \delta {\cal E}_0, 
\end{align}
where ${\cal E}_0$ is the vacuum energy for $\mu_3 = 0$, i.e., ${\cal E}_0 = \mu_1/2\lambda_1$, 
and from Eq.~(\ref{eq:pot3}) we have
\begin{align}
\delta {\cal E}_0 = 
C\, \mu_3^2 + {\cal O}(\mu_3^3),
\end{align}
in which $C$ is a coefficient determined by other coupling constants.

It is now clear that $\mu_3 = 0$ corresponds to the extremum of ${\cal E}$, and such a point is naturally realized by the MPP in the large volume limit. 
(See the discussion of the MPP in the Introduction.)
Note that we have not assumed any fine-tuning for other couplings, and our conclusions hold for a wide range of parameter space satisfying inequalities such as (\ref{eq:condition2}). 

\section{Towards Yukawa alignment via MPP} \label{sec:Yukawa-alignment}

\begin{figure}[t]
\begin{center}
\includegraphics[width=150mm]{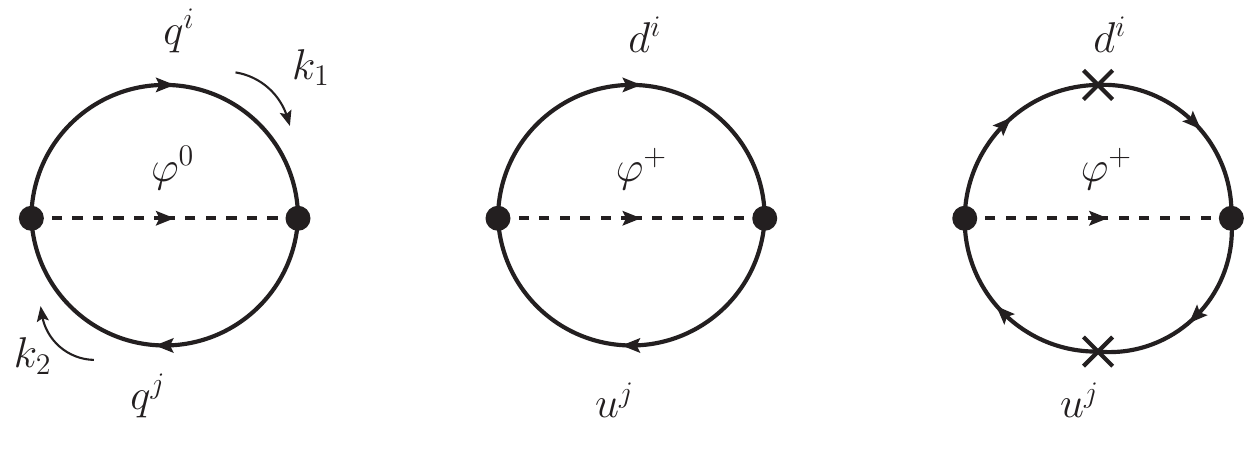}
\caption{Vacuum bubble diagrams with quark loops. }
\label{fig:yukawa}
\end{center}
\end{figure}

We discuss a possible mechanism that may realize the Yukawa alignment in the 2HDM.
We impose the Higgs alignment which can naturally be realized by considering the MPP as we have seen in the previous section. 
In this case, flavor-changing Yukawa couplings arise from those for $\varphi^\pm$ and $\varphi^0$, see Eq.~(\ref{BottomUp:2HDM:Yukawa1}). 

Let us consider the vacuum bubble diagrams shown in Fig.~\ref{fig:yukawa}, focusing on the quark loops as an example. The amplitude is calculated as
\begin{align}
{\cal E}_{\rm bubble} 
 & = -\int \frac{d^4k_1}{(2\pi)^4}\int \frac{d^4k_2}{(2\pi)^4}\frac{1}{(k_1-k_2)^2 - m_{\varphi}^2} \notag\\
&\qquad\times\Bigg\{\left(2k_1\cdot k_2\right)\text{tr} \Bigg[
 \sum_{q = u,d}\left(\frac{1}{k_1^2 - M_q^2}\rho_q \frac{1}{k_2^2 - M_q^2} \rho_q^\dagger \right) \notag\\
&\phantom{\qquad\times\Bigg\{\left(2k_1\cdot k_2\right)\text{tr} \Bigg[} +  \frac{1}{k_1^2 - M_u^2}V_{\rm CKM}\rho_d \frac{1}{k_2^2 - M_d^2} \rho_d^\dagger V_{\rm CKM}^\dagger \notag\\
&\phantom{\qquad\times\Bigg\{\left(2k_1\cdot k_2\right)\text{tr} \Bigg[}
    + \frac{1}{k_1^2 - M_u^2}\rho_u^\dagger V_{\rm CKM} \frac{1}{k_2^2 - M_d^2} V_{\rm CKM}^\dagger \rho_u
\Bigg]\notag\\
&
\phantom{\times\Bigg\{\left(2k_1\cdot k_2\right)\text{tr} \Bigg[}
\hspace{-3mm}    -\text{tr} \Bigg[\frac{M_u}{k_1^2 - M_u^2}V_{\rm CKM}\rho_d \frac{M_d}{k_2^2 - M_d^2} V_{\rm CKM}^\dagger \rho_u  \notag\\
&\phantom{\qquad\times\Bigg\{\left(2k_1\cdot k_2\right)\text{tr} \Bigg[}+ \frac{M_u}{k_1^2 - M_u^2}\rho_u^\dagger V_{\rm CKM} \frac{M_d}{k_2^2 - M_d^2} \rho_d^\dagger V_{\rm CKM}^\dagger\Bigg]
\Bigg\}, \label{eq:e_bubble}
\end{align}
where terms proportional to $(2k_1\cdot k_2)$ arise from the left and center diagrams, while the contribution in the last two lines originate from the right diagram in Fig.~\ref{fig:yukawa}. For simplicity, we assume the masses of neutral and charged scalars to be identical, denoted as $m_{\varphi}$.

Let us consider the contribution from the exchange of the neutral Higgs boson $\varphi^0$ (left diagram in Fig.~\ref{fig:yukawa}), corresponding to the first term of the second line in Eq.~(\ref{eq:e_bubble}). After taking the trace in the flavor space, we can express this contribution by terms proportional to $|\rho_q^{\rm diag}|^2$ and $|\delta \rho_q|^2$, where $\rho_q^{\rm diag}$ and $\delta \rho_q$ represent the diagonal and off-diagonal elements of $\rho_q$, respectively. Consequently, ${\cal E}_{\rm bubble}$ reaches an extremum at $\delta \rho_q = 0$ in the parameter space, and the diagonal form of $\rho_q$ is naturally realized by the MPP:
\begin{align}
\frac{\partial {\cal E}_{\rm bubble}(\rho_q)}{\partial \rho_q^{ij}} = 0
\quad(i \neq j)
\qquad\text{at}\quad\rho_q = \rho_q^{\rm diag}. \label{eq:mpp-yukawa}
\end{align}
However, this description is modified by the contribution from the exchange of charged Higgs~$\varphi^\pm$, as $\rho_q$ is coupled with the CKM matrix $V_{\rm CKM}$. Therefore, the extrema discussed above are shifted by the off-diagonal elements of $V_{\rm CKM}$.

We would like to comment on the lepton sector. In this context, the quark states $q^i$ in the left diagram are replaced by the charged lepton states $e^i$, and the pairs $(d^i,u^j)$ in the center and right diagrams are replaced by $(e^i,\nu^j)$. By neglecting neutrino masses, the vacuum energy ${\cal E}_{\rm bubble}$ can be calculated by substituting $(\rho_d,\rho_u,M_d,M_u,V_{\rm CKM})$ with $(\rho_e,0,M_e,0,I_{3\times 3})$ in Eq.~(\ref{eq:e_bubble}).\footnote{When we introduce three right-handed neutrinos, another Yukawa matrix $\rho_\nu$ appears in Eq.~(\ref{BottomUp:2HDM:Yukawa1}), similar to $\rho_u$. Even in this case, the MPP predicts a diagonal form of $\rho_\nu$ as long as we take the massless limit of the neutrinos. } 
Consequently, Eq.~(\ref{eq:e_bubble}) is satisfied, and the diagonal form of $\rho_e$ is naturally realized by the MPP.

\section{Conclusion \label{sec:conclusion}}

The general two Higgs doublet model (2HDM) encounters fine-tuning problems to cope with current experimental data, specifically regarding the realization of the Higgs and Yukawa alignments. In this work, we have implemented the multicritical point principle (MPP) to solve these problems.

We have shown in the basis with 
$\lambda_6 = 0$ (a dimensionless parameter in the Higgs potential) that setting the dimensionful mixing parameter $\mu_3$ between the two doublets to zero leads to the vanishing vacuum expectation value (VEV) of the second Higgs doublet in a finite region of the parameter space, bounded by several inequalities. 
Moreover, we have demonstrated that the point $\mu_3 \rightarrow 0$ coincides with a critical point—specifically an extremum of the vacuum energy. Thus, the Higgs alignment is naturally deduced through the MPP.

%

For the Yukawa alignment, our analysis shows that the MPP effectively predicts the Yukawa alignment in the quark sector
assuming negligible off-diagonal elements of the Cabibbo-Kobayashi-Maskawa matrix. 
The predicted Yukawa matrices $\rho_f$ for the additional Higgs bosons present a more general structure unlike the one given by the conventional Yukawa alignment, i.e., 
these matrices can have three different values in their diagonal complex elements.
The Yukawa alignment for the lepton sector is naturally deduced by the MPP assuming massless neutrinos.  


This study's outcomes not only deepen our understanding of the Higgs mechanism but also pave the way for experimental tests of the MPP in 2HDM settings. While predicated on certain assumptions, the scope of this research potentially reaches into broader realms, such as cosmology and the underlying principles of particle physics. Future advancements in technology and cross-disciplinary research could further elucidate these findings and their broader impact.

\begin{acknowledgments}

H.K.\ thanks Prof. Shin-Nan Yang and his family for their kind support through the Chin-Yu chair professorship. H.K. is partially supported by JSPS (Grants-in-Aid for Scientific Research Grants No.~20K03970), by the Ministry of Science and Technology, R.O.C. (MOST 111-2811-M-002-016), and by National Taiwan University.
The work of K.K.\ 
is supported by KIAS Individual Grants, Grant No. 090901, and
that of K.O.\ is in part supported by the JSPS Kakenhi Grant No.~21H01107.

\end{acknowledgments}

\begin{appendix}

\section{Canonical form of the kinetic term \label{sec:canonical}}  
The basis~(\ref{eq:kawai-basis}) results in a non-canonical kinetic term for massive gauge bosons, owing to the mixing between the gauge boson and the corresponding Nambu-Goldstone boson. We will detail the process of achieving a canonical kinetic term in this context. To simplify our discussion, we assume that $\langle \varphi^\pm \rangle = 0$.

It is convenient to decompose the massive gauge bosons as
\begin{align}
W_\mu^\pm = \tilde{W}_\mu^\pm +  \frac{1}{m_W^{}}\partial_\mu\phi_W^\pm,  \quad
Z_\mu = \tilde{Z}_\mu + \frac{1}{m_Z^{}}\partial_\mu\phi_Z, 
\end{align}
where $\tilde{W}_\mu^\pm$ ($\tilde{Z}_\mu$) 
and $\phi_W^\pm$ ($\phi_Z^{}$) are respectively the 
transverse and longitudinal modes of the $W_\mu^\pm$ ($Z_\mu)$ boson. 
We note that $\partial^\mu\tilde{W}_\mu^\pm = \partial^\mu\tilde{Z}_\mu = 0 $ are 
satisfied by definition. 
The kinetic term is then expressed as 
\begin{align}
{\cal L}_{\rm kin} & = (\partial_\mu \phi_W^-, \partial_\mu \varphi^-)
\begin{pmatrix}
1 & c^*\\
c & 1
\end{pmatrix}
\begin{pmatrix}
\partial^\mu \phi_W^+\\ 
\partial^\mu \varphi^+
\end{pmatrix} \notag\\
&\quad+\frac{1}{2}
(\partial_\mu \phi_Z, \partial_\mu \varphi_\I ^{}, \partial_\mu \varphi_\R ^{})
\begin{pmatrix}
1 & \frac{v_\R ^{}}{v} & -\frac{v_\I ^{}}{v} \\
\frac{v_\R ^{}}{v} & 1 & 0 \\
-\frac{v_\I ^{}}{v} & 0 & 1
\end{pmatrix}
\begin{pmatrix}
\partial_\mu \phi_Z\\
\partial_\mu \varphi_\I ^{}\\
\partial_\mu \varphi_\R ^{}
\end{pmatrix}
,\quad c = -i\frac{v_\R ^{} + i v_\I ^{}}{v}, 
\end{align}
where $v \equiv \sqrt{v_\eta^2 + v_\R ^2 + v_\I ^2}$ with $v_\eta = \langle \eta \rangle$, 
$v_\R ^{} = \langle \varphi_\R  \rangle$ and 
$v_\I ^{} = \langle \varphi_\I  \rangle$. 
The kinetic terms for the transverse modes $\tilde{W}_\mu^\pm$ and $\tilde{Z}_\mu$
 already take the canonical form. 
By introducing the non-unitary transformation
\begin{align}
\begin{pmatrix}
\phi_W^+ \\
\varphi^+
\end{pmatrix}
=
\begin{pmatrix}
\frac{c^*}{|c|} & - \frac{c^*}{\sqrt{1 - |c|^2}} \\
0             & \frac{1}{\sqrt{1 - |c|^2}}
\end{pmatrix}
\begin{pmatrix}
\phi_W^{+\prime} \\
\varphi^{+\prime}
\end{pmatrix}, ~~
\begin{pmatrix}
\phi_Z\\
\varphi_\I ^{}\\
\varphi_\R ^{}
\end{pmatrix}
=
\begin{pmatrix}
 1 & -\frac{\rho }{\sqrt{1-\rho ^2}} & 0 \\
 0 & \frac{\rho_\R }{\rho  \sqrt{1-\rho^2}} & \frac{\rho_\I }{\rho} \\
 0 & -\frac{\rho_\I }{\rho  \sqrt{1-\rho ^2} }& \frac{\rho_\R }{\rho } 
\end{pmatrix}
\begin{pmatrix}
\phi_Z^\prime\\
\varphi_\I ^\prime\\
\varphi_\R ^\prime
\end{pmatrix}, 
\end{align}
$\phi_W^{\pm\prime}$ and $\varphi^{\pm\prime}$ can take the canonical form of the kinetic term, and we can identify the former (latter)
with the unphysical NG boson (physical charged Higgs boson), where 
$\rho = \sqrt{v_\R ^2 + v_\I ^2}/v$ and $\rho_{\R,\I}^{} = v_{\R,\I}^{}/v$.

 \end{appendix}

\bibliography{references}
\end{document}